%% file: pscc2026_template.tex
\documentclass{IEEEtran4PSCC}
\ifCLASSINFOpdf
   \usepackage[pdftex]{graphicx}
\else
   \usepackage[dvips]{graphicx}
\fi
\usepackage[cmex10]{amsmath}
\usepackage{amssymb}
\usepackage{xcolor}
\usepackage{tikz}
\usepackage{comment}
\usepackage[hyphens]{url}
\usepackage{hyperref}
\usepackage{breakurl}

\usepackage{dsfont}
\begin{document}
%
\title{Bayesian Inference for Estimating Generation Costs in Electricity Markets}
\input{./Authors}



\maketitle

\begin{abstract}
Estimating generation costs from observed electricity market data is essential for market simulation, strategic bidding, and system planning. To that end, we model the relationship between generation costs and production schedules with a latent variable model. Estimating generation costs from observed schedules is then formulated as Bayesian inference. A prior distribution encodes an initial belief on parameters, and the inference consists of updating the belief with the posterior distribution given observations. We use balanced neural posterior estimation (BNPE) to learn this posterior. Validation on the IEEE RTS-96 test system shows that marginal costs are recovered with narrow credible intervals, while start-up costs remain largely unidentifiable from schedules alone. The method is benchmarked against an inverse-optimization algorithm that exhibits larger parameter errors without uncertainty quantification.
\end{abstract}


\section{Introduction}
\label{sec:intro}
Electricity is exchanged across multiple markets by many agents, from years to minutes before actual delivery \cite{wood2013power, kirschen2018fundamentals}. Agents make decisions in this complex process based on incomplete or imperfect information. Estimation of generation costs (marginal and start-up) from observed market data (generation schedule) enables more accurate price forecasts, risk assessment, and long-term revenue estimation, benefiting analysts in utilities and hedge funds \cite{zareipour2009economic,JRC2025Forecasting}. Such estimates further allow regulators and TSOs to monitor market behavior and evaluate capacity retention in zonal markets where localized price signals are limited.

Producers aim to maximize profit by selling production through electricity markets. In a perfect market, participants bid at their marginal costs for all future periods available on that market, i.e., they propose to sell electricity at the cost of producing each additional MWh. In theory, as time advances, bids evolve as new information becomes available until the delivery period. In practice, several electricity markets exist to approximate this ideal behavior. We focus on an ideal day-ahead market, where production schedules for the day are fixed one day in advance for each participant. We assume that each participant submits independent bids for each production unit. Equivalently, each unit corresponds to a unique participant.

A unit commitment (UC) problem \cite{padhy2004unit} is standardly solved to approximate the market clearing process. This mixed-integer optimization problem minimizes total generation cost while satisfying physical constraints such as capacity limits, ramp rates, minimum up/down periods, and transmission limits. Large-scale UC problems can be solved efficiently using modern formulations \cite{carrion2006computationally, gentile2017tight} and commercial solvers. Using UC to approximate market clearing is justified by the assumption that in a competitive environment, the collective result of participant bidding aligns with the cost-optimal physical dispatch of the system's assets \cite{mas1995microeconomic, hobbs2001linear, wilson2002architecture}. Consequently, market participants and system operators use UC models as a proxy to simulate market outcomes and forecast prices and volumes.

Literature on UC divides into two main categories. Forward approaches treat cost parameters as known and seek to determine generation schedules. Operational uncertainties are addressed through stochastic programming \cite{birge1997introduction, takriti1996stochastic, zheng2014stochastic, haaberg2019fundamentals} or robust optimization \cite{bertsimas2012adaptive}. Inverse approaches treat cost parameters as unknown and seek to infer them from observed schedules. They typically consist of solving an inverse optimization problem to recover parameters \cite{ahuja2001inverse, birge2017inverse}. For example, Liang and Dvorkin \cite{liang2023data} apply such a method to estimate bid prices in US nodal markets from locational marginal prices and generator schedules. This approach relies on bus-level price information unavailable in zonal markets such as those in Europe \cite{meeus2005development}. Inverse optimization methods often return point estimates and provide limited uncertainty quantification on the recovered parameters. Preliminary work addresses this limitation by applying simulation-based inference \cite{cranmer2020frontier} to generation cost parameter estimation in UC problems \cite{pirlet2024cost}. It demonstrates the feasibility of using neural posterior estimation (NPE) \cite{rezende2015variational, papamakarios2016fast, lueckmann2017flexible, greenberg2019automatic}  to estimate generation cost in a simplified 9-unit system without network constraints. The present work extends this foundation to a realistic network and compares to a traditional inversion approach.

In this work, we extend the UC formulation to model stochastic demand, line and generation outages, and strategic bidding, all of which are reflected in market schedules. To that end, we use a latent variable model and formulate cost estimation as a Bayesian inference problem. This allows quantifying uncertainty about generation costs and updating estimates as data is observed. We use balanced neural posterior estimation (BNPE) to approximate posterior distributions, enabling fast repeated inference with uncertainty quantification. We also introduce an inverse-optimization baseline that provides fast point estimates when full uncertainty quantification is not required. We validate both approaches on the IEEE RTS-96 test system \cite{grigg1999ieee} with its 2016 update \cite{ordoudis2016updated}. Methods are compared based on estimation accuracy, computational efficiency, and uncertainty characterization.

The remainder of the paper is organized as follows. In Section~\ref{sec:prob_statement}, the market model and the Bayesian inference problem for recovering generation costs are formulated. The two practical algorithms are described in Section~\ref{sec:methodology} and validated in Section~\ref{sec:experiments}. Finally, conclusions are highlighted in Section~\ref{sec:conclusion}.


\section{Problem Statement}
\label{sec:prob_statement}

In this section, we first define a latent variable model \cite{blei2014build} that describes how electricity market schedules are obtained depending on generation costs. We then formalize the problem of estimating costs as a Bayesian inference problem with the previous model.

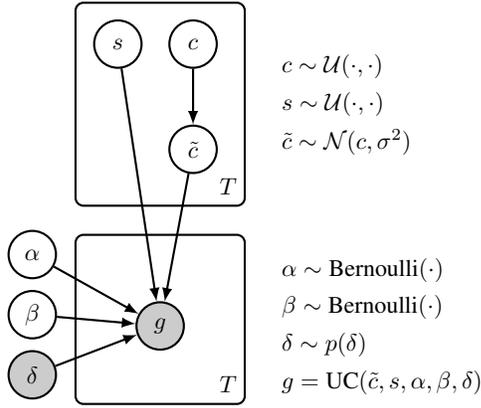
\begin{figure}[h]
\centering
\begin{tikzpicture}[>=latex, thick, auto, scale=1, every node/.style={scale=0.9}]
\tikzstyle{var}=[circle, draw=black, minimum size=0.7cm, inner sep=0pt]
\tikzstyle{box}=[rectangle, draw=black, rounded corners=3pt, inner sep=5pt]
\tikzstyle{obs}=[circle, draw=black, fill=gray!40, minimum size=0.7cm, inner sep=0pt]

\node[box, minimum width=2.5cm, minimum height=3cm] (upper) {};
\node[var] (s) at ([xshift=-1.7cm,yshift=0.8cm]upper.east) {$s$};
\node[var] (c) at ([xshift=-0.7cm,yshift=0.8cm]upper.east) {$c$};
\node[var] (ctilde) at ([xshift=-0.7cm, yshift=-0.6cm]upper.east) {$\tilde{c}$};

\node[anchor=north east] (t) at ([yshift=0.5cm]upper.south east) {$T$};

\draw[->] (c) -- (ctilde);

\node[box, minimum width=2.5cm, minimum height=2.5cm] (lower) at ([yshift=-1.5cm]upper.south) {};
\node[obs] (g) at (0, -2.95cm) {$g$};

\node[var] (alpha) at (-1.7cm, -2cm) {$\alpha$};
\node[var] (beta) at (-1.7cm, -2.8cm) {$\beta$};
\node[obs] (delta) at (-1.7cm, -3.6cm) {$\delta$};

\node[anchor=north east] (t) at ([yshift=0.5cm]lower.south east) {$T$};

\draw[->] (alpha) -- (g);
\draw[->] (beta) -- (g);
\draw[->] (delta) -- (g);

\draw[->] (s) -- (g);
\draw[->] (ctilde) -- (g);

\node[anchor=west] at (1.5cm, 0.5cm) {$c \sim \mathcal{U}(\cdot, \cdot)$};
\node[anchor=west] at (1.5cm, 0cm) {$s \sim \mathcal{U}(\cdot, \cdot)$};
\node[anchor=west] at (1.5cm, -0.5cm) {$\tilde{c} \sim \mathcal{N}(c, \sigma^2)$};

\node[anchor=west] at (1.5cm, -2.2cm) {$\alpha \sim \text{Bernoulli}(\cdot)$};
\node[anchor=west] at (1.5cm, -2.7cm) {$\beta \sim \text{Bernoulli}(\cdot)$};
\node[anchor=west] at (1.5cm, -3.2cm) {$\delta \sim p(\delta)$};
\node[anchor=west] at (1.5cm, -3.7cm) {$g = \text{UC}(\tilde{c}, s, \alpha, \beta, \delta)$};
\end{tikzpicture}
\caption{Latent variable model for cost inference. White circles represent latent variables, gray circles represent observed variables. The two rectangular boxes denote the $T$ time periods over which the model operates, with variables influencing each other across time. In the upper box are marginal costs $c$ (latent, uniform prior) and start-up costs $s$ (latent, uniform prior). Marginal costs pass through a stochastic bidding model $\tilde{c} \sim \mathcal{N}(c, \sigma^2)$ (latent), while start-up costs remain unchanged. In the lower part are generator availability $\alpha$ (latent, Bernoulli), line availability $\beta$ (latent, Bernoulli), observed system load $\delta$ (gray circle). The deterministic UC optimization problem combines bidding costs $\tilde{c}$, start-up costs $s$, availabilities $\alpha, \beta$, and observed system load $\delta$ (gray circle) sampled from a distribution capturing seasonal, weekly, and diurnal demand patterns, to produce the schedule $g$ (observed, gray circle).}
\label{fig:uc_flow}
\end{figure}

The latent variable model describing the market interactions is illustrated in Figure~\ref{fig:uc_flow}. Formally, let us describe the electrical network with a graph composed of nodes $\mathcal{N}$ and edges $\mathcal{E} \subset \mathcal{N} \times \mathcal{N}$. Let $\mathcal{T}=\{1,\dots,T\}$ index time periods and $\mathcal{J}=\{1,\dots,J\}$ index generators. The model parameters include the marginal generation cost $c_{tj}$ and the start-up cost $s_{tj}$ for each generator $j \in \mathcal{J}$ at time $t \in \mathcal{T}$. We consider uniform priors on these parameters $c_{tj} \sim \mathcal{U}(c_{\min}, c_{\max})$ and $s_{tj} \sim \mathcal{U}(s_{\min}, s_{\max})$, which model uncertainty over a range of feasible values. The price offered for producing energy depends on the marginal costs and accounts for trading strategies, which is modeled with the bidding costs $\tilde{c}_{tj} \sim \mathcal{N}(c_{tj}, \sigma^2)$. The availability of generator $j \in \mathcal{J}$ is modeled with the Bernoulli random variable $\alpha_{j} \sim \text{Bernoulli}(1 - p_{\text{gen}})$ and the availability of the line $e \in \mathcal{E}$ with $\beta_{e} \sim \text{Bernoulli}(1 - p_{\text{line}})$. The demand at bus $n$ and time $t$ is modeled as $\delta_{tn} = L_t \cdot s_{tn}$, where $L_t$ is the total system load following seasonal and diurnal profiles specified in the IEEE RTS-96 test system and $s_{tn}$ are the bus-level load shares, normalized such that $\sum_{n} s_{tn} = 1$. Both variables are disturbed by a Gaussian noise before scaling such that the demand vector $\delta$ follows the distribution $p(\delta)$. Generation schedules $g$ are the solution of the UC problem, which minimizes total generation cost subject to physical and operational constraints such as capacity limits, ramp rates, minimum up/down times, and transmission limits. The generation schedule depends on the realizations of the previous events
\begin{align}
    g = UC(\tilde{c}, s, \alpha, \beta, \delta).
\end{align}

We denote the market outcome by $x = (g, \delta)$, the cost parameters by $\theta=(c, s)$ and the latent variables by $z=(\tilde{c}, \alpha, \beta)$. Finally, the forward model is defined by the joint distribution of these variables
\begin{align}
    p(x, z, \theta) = p(g \mid z, \theta, \delta) \; p(z \mid \theta) \;p(\theta) \; p(\delta), \label{eq:joint}
\end{align}
where $p(z \mid \theta) = p(\tilde{c} \mid c)\, p(\alpha)\, p(\beta)$ and $p(g \mid z, \theta, \delta)$ is a Dirac centered in the solution of the UC.

The problem of finding cost parameters $\theta$ from observed market outcome $x^{\mathrm{obs}} = (g^{\mathrm{obs}}, \delta^{\mathrm{obs}})$ is formulated as a Bayesian inference problem. We represent our knowledge on plausible cost parameters before observing market outcome using the prior distribution $p(\theta)$, and apply Bayes' theorem to represent plausible cost parameters after observation using the posterior distribution 
\begin{equation}
    \label{eq:bayesian_problem}
    p(\theta \mid x^{\mathrm{obs}}) \propto p(\theta) \, p(x^{\mathrm{obs}} \mid \theta) \,.
\end{equation}


\section{Methodology} \label{sec:methodology}

To solve the Bayesian inference problem defined in Eq.~\eqref{eq:bayesian_problem}, we present balanced neural posterior estimation (BNPE) \cite{delaunoy2023balancing}, a simulation-based inference method, to approximate the posterior distribution from simulations. We compare this approach with a baseline method using inverse optimization, which provides fast point estimates without quantifying uncertainty. These two methods offer different trade-offs between estimation accuracy, computational efficiency, and uncertainty characterization.

\subsection{Simulation-based posterior inference with BNPE}
\label{subsec:bnpe}

Computing the posterior distribution from Bayes' theorem would require integrating over latent variables, which is intractable in practice. Instead, we use BNPE, in which we train a neural network $q_\phi(\theta \mid x)$, where $\phi$ denotes the weights of the network, to approximate the posterior $p(\theta \mid x)$. This network is trained by minimizing the expected Kullback-Leibler (KL) divergence between the true posterior and the neural network approximation
$$\min_{\phi} \underset{p(x)}{\mathbb{E}} \left[ \mathrm{KL} \big( p(\theta \mid x) \parallel q_\phi(\theta \mid x) \big) \right],$$
where the expectation is taken over the marginal distribution of simulated market outcomes $p(x) = \int p(x \mid \theta) p(\theta) d\theta$.

The KL divergence measures the information loss when approximating $p(\theta \mid x)$ with $q_\phi(\theta \mid x)$. As shown in Appendix~\ref{app:bnpe_derivation}, this is equivalent to minimizing the expected negative log-posterior
$$\min_{\phi} \underset{p(\theta, x)}{\mathbb{E}} \left[ - \log q_\phi(\theta \mid x) \right],$$
where $p(\theta, x) = p(x \mid \theta) p(\theta)$ denotes the joint distribution induced by the latent variable model.

Training consists in generating samples $(\theta, x)$ by sampling $\theta$ from the prior $p(\theta)$, sampling $\delta$ from the load distribution $p(\delta)$, and computing the schedule $g$ through the UC optimization, resulting in $x = (g, \delta)$. These samples are used to estimate the expected log-posterior, which is minimized via stochastic gradient descent on the network parameters $\phi$. Once trained, the density $q_{\phi}(\theta \mid x^{\mathrm{obs}})$ serves as a surrogate for $p(\theta \mid x^{\mathrm{obs}})$ when observing market outcome $x^{\mathrm{obs}}$.


\subsection{Inverse optimization with polar cone method}

We compare our approach with inverse optimization \cite{birge2017inverse}. This method seeks cost parameters for which the observed schedule is optimal under a deterministic model. We therefore neglect stochastic bidding and random outages to obtain a deterministic model, namely the UC problem and
\begin{equation}
    \label{eq:UC_deterministic}
    g = \text{UC}(c, s, \mathbf{1}, \mathbf{1}, \delta).
\end{equation}

Let us first note that for a given observed load $\delta^{\mathrm{obs}}$, the observed schedule $g^{\mathrm{obs}}$ is optimal when solving the UC problem with any parameter $\theta$ that satisfies the condition
\begin{equation}
    \label{eq:optim_criterion}
    \theta^\top(g - g^{\mathrm{obs}}) \geq 0 \quad \forall g \in \mathcal{G}(\delta^{\mathrm{obs}}),
\end{equation}
where $\mathcal{G}(\delta^{\mathrm{obs}})$ is the set of feasible schedules in the UC problem given the observed load $\delta^{\mathrm{obs}}$. The set of parameters that satisfy this condition is denoted by $\Theta^\star$, which is usually called the polar cone \cite{rockafellar1970convex}.

The inverse optimization problem refines iteratively an approximation of the polar cone. The initial approximation is the hypercube $\Theta^{(0)} = [\theta^{\text{min}}, \theta^{\text{max}}]$ representing physically plausible cost ranges. At each iteration $k$, we sample a reference parameter $\theta^{\text{ref}}$ from the prior that we project onto the current approximation $\Theta^{(k)}$ solving
\begin{align}
    \label{eq:projection}
    \min_{\theta \in \Theta^{(k)}} \quad & \|\theta - \theta^{\text{ref}}\|.
\end{align}
Let $\theta^{(k)}$ denote the solution of the projection. We solve the UC problem from Eq.~\eqref{eq:UC_deterministic} with these costs to obtain the schedule $g^{(k)}$ and verify the optimality condition~\eqref{eq:optim_criterion}. If $\theta^{(k)\top}(g^{(k)} - g^{\mathrm{obs}}) = 0$, then the observed schedule is optimal for the current cost parameters and the algorithm has converged. If $\theta^{(k)\top}(g^{(k)} - g^{\mathrm{obs}}) < 0$, then $\theta^{(k)}$ violates condition~\eqref{eq:optim_criterion}, and we refine the polar cone approximation $\Theta^{(k+1)} = \{ \theta \in \Theta^{(k)} : \theta^{\top}(g^{(k)} - g^{\mathrm{obs}}) \geq 0 \}$.

As the number of iterations increases, $\Theta^{(k)}$ converges to the polar cone $\Theta^\star$, and $g^{\mathrm{obs}}$ is an optimal schedule for any solution of the projection~\eqref{eq:projection}. In practice, we stop when $\theta^{(k)\top}(g^{(k)} - g^{\mathrm{obs}}) \geq -\varepsilon$, meaning that the observed schedule is $\varepsilon$-optimal, and the parameter $\theta^{(k)}$ serves as the approximate solution.

This method is only applicable for inverting deterministic models. We therefore ignore stochastic random bidding and model outages, which introduces model misspecification. Despite this limitation, inverse optimization provides fast point estimates for time-critical applications.


\section{Experiments} \label{sec:experiments}

\subsection{Experimental setup}

We apply our two algorithms on the IEEE RTS-96 test system \cite{grigg1999ieee} with its 2016 update \cite{ordoudis2016updated}. This system includes 24 buses $n \in \mathcal{N}$ (17 buses with loads, and $J=12$ generators connected to 10 buses), and 38 transmission lines $ e \in \mathcal{E}$. This problem covers a time horizon of $T = 24$ hours.

We estimate cost parameters for each generator, i.e., marginal costs $c_{tj}$ and start-up costs $s_{tj}$. We assume these costs remain constant over the 24-hour horizon such that
\begin{align*}
    c_{tj} = c_j, \quad \forall t \in \mathcal{T}\\
    s_{tj} = s_j. \quad \forall t \in \mathcal{T} 
\end{align*}

We define uniform priors for marginal costs $c_j\sim \mathcal{U}(10, 50)$€/MWh and start-up costs $s_j \sim \mathcal{U}(500, 10000)$€. These ranges reflect typical values observed in electricity markets while encoding minimal prior knowledge.

We model bidding costs with the transition $\tilde{c}_j \sim \mathcal{N}(c_j, \sigma^2)$ with $\sigma = 2.5$ €/MWh. We model generator and transmission line availability with $\alpha_j \sim \text{Bernoulli}(0.95)$ and $\beta_{ij} \sim \text{Bernoulli}(0.99)$, respectively.

Daily demand scenarios are generated by applying the modeling described in Section~\ref{sec:prob_statement} to the IEEE RTS-96 system. Specifically, we set the total load noise to $\sigma_L = 5\%$ and the bus-share perturbation to $\sigma_s = 0.5\%$, using the seasonal and diurnal profiles from \cite{ordoudis2016updated} as the base temporal patterns.

The UC model is a security-constrained unit commitment (SCUC) \cite{yang2021comprehensive} model, which minimizes total generation cost subject to generator operational constraints (capacity limits, ramp rates, minimum up/down times) and transmission network constraints (thermal limits of line under DC power flow). The complete mathematical formulation is as follows.

\textit{Technical parameters $\psi$}
\begin{align}
    &G_j^{\min},G_j^{\max} \quad\text{min/max generation of unit }j, \notag \\
    &ru_j,rd_j \quad\text{ramp-up / ramp-down limits for unit }j, \notag\\
    &mu_j,md_j \quad\text{minimum up / down times (integer) for unit }j, \notag \\
    &B_{ij} \quad \text{susceptance of line }(i,j), \notag \\
    &\overline{F}_{ij} \quad \text{thermal capacity if line }(i,j), \notag \\
    &v^{\mathrm{init}}_j,g^{\mathrm{init}}_j \quad \text{initial on/off status and power of unit }j, \notag \\
    &\alpha_j \quad \text{availability indicators for generator }j \notag,\\
    &\beta_{ij} \quad \text{availability indicators for line }(i,j), \notag \\
    &n_{\text{slack}} \in \mathcal{N} \quad \text{index of the slack (reference) bus.} \notag \\
\end{align}
\textit{System load}
\begin{align}
    &\delta_{tn} \quad \text{demand at bus }n\text{ and time }t \notag.
\end{align}
\textit{Cost parameters $\theta = [c, s]$} (unknown in the inverse problem) \begin{align} 
    &c_j \quad \text{marginal cost of unit }j, \notag \\
    &s_j \quad \text{start-up cost of unit }j \notag.
\end{align}
\textit{Decision variables}
\begin{align}
  &g_{tj}\in\mathbb{R}_+ \quad \text{dispatch of unit }j\text{ at time }t, \notag\\
  &v_{tj}\in\{0,1\} \quad \text{commitment (on/off) of unit }j\text{ at time }t, \notag\\
  &y_{tj}\in\{0,1\} \quad \text{start-up indicator of unit }j\text{ at time }t, \notag\\
  &z_{tj}\in\{0,1\} \quad \text{shut-down indicator of unit }j\text{ at time }t, \notag\\
  &\vartheta_{tn}\in\mathbb{R} \quad \text{voltage angle at bus }n\text{ and time }t, \notag\\
  &f_{t \, ij}\in\mathbb{R} \quad \text{directed flow on line }(i,j)\text{ at time }t \notag.
\end{align}
\textit{Constraints}\\
\textit{Slack bus}
$$\vartheta_{t \,n_{\text{slack}}}=0\qquad\forall t\in\mathcal{T}.$$
\textit{DC power flow and thermal limits:}
\begin{align}
    f_{t\, ij}&=B_{ij}\big(\vartheta_{ti}-\vartheta_{tj}\big), \notag \\
    -\beta_{ij}\,\overline{F}_{ij}\le &f_{t \,ij}\le \beta_{ij}\,\overline{F}_{ij},\quad\forall (i,j)\in\mathcal{E},\ \forall t\in\mathcal{T} \notag.
\end{align}
\textit{Generation bounds:}
$$\alpha_j G_j^{\min}\, v_{tj}\ \le\ g_{tj}\ \le\ \alpha_j G_j^{\max}\, v_{tj}
\quad\forall j\in\mathcal{J},\ \forall t\in\mathcal{T}.$$
\textit{State-transition (start/stop):}
$$y_{tj}-z_{tj}= 
\begin{cases}
v_{1j}-v^{\mathrm{init}}_j & t=1,\\[4pt]
v_{tj}-v_{t-1 \,j} & t > 1,
\end{cases}
\quad\forall j\in\mathcal{J}.$$
\textit{Ramp constraints for} $t=1$:
\begin{align}
&g_{1j}-g^{\mathrm{init}}_j \le ru_j\,v^{\mathrm{init}}_j + G_j^{\min}\,y_{1j}, \quad\forall j\in\mathcal{J},\notag\\
&g^{\mathrm{init}}_j - g_{1j} \le rd_j\,v^{\mathrm{init}}_j + G_j^{\min}\,z_{1j}, \quad\forall j\in\mathcal{J}\notag.
\end{align}
\textit{Ramp constraints for} $t > 1$:
\begin{align}
&g_{tj}-g_{t-1\, j} \le ru_j\,v_{t-1 \, j} + G_j^{\min}\,y_{tj}, \quad\forall j\in\mathcal{J}, \notag\\
&g_{t-1 \, j}-g_{tj} \le rd_j\,v_{tj} + G_j^{\min}\,z_{tj}, \quad\forall j\in\mathcal{J} \notag.
\end{align}
\textit{Minimum up/down time constraints:}\\
For $t\ge mu_j$
$$\sum_{k=t-mu_j+1}^{t} y_{kj}\ \le\ v_{tj}, \quad\forall j\in\mathcal{J},$$
and for $t\ge md_j$
$$\sum_{k=t-md_j+1}^{t} z_{kj}\ \le\ 1-v_{tj} \quad\forall j\in\mathcal{J}.$$
\textit{Nodal power balance:}
For each bus $n\in\mathcal{N}$ and time $t\in\mathcal{T}$
$$\sum_{j\in\mathcal{J}_n} g_{tj} - \delta_{tn} = \sum_{m:\,(n,m)\in\mathcal{L}} f_{t \,n m} -
\sum_{m:\,(m,n)\in\mathcal{L}} f_{t\, m n},$$
where $\mathcal{J}_n\subseteq\mathcal{J}$ is the set of generators connected to bus $n$.
\textit{Objective (cost minimization)}\\
$$\min_{g,v,y,z,\vartheta,f}\; 
\sum_{t\in\mathcal{T}}\sum_{j\in\mathcal{J}}\big(c_j\,g_{tj} + s_j\,y_{tj}\big).$$

\begin{figure}
    \centering
    \includegraphics[width=\linewidth]{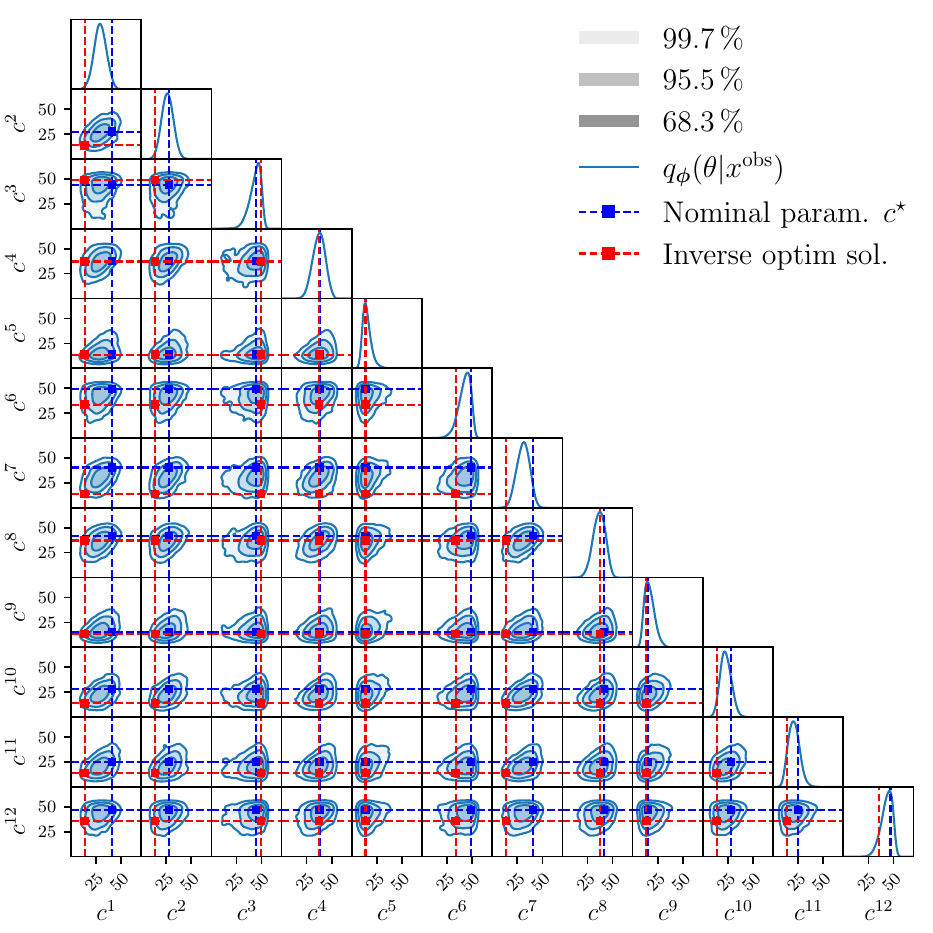}
    \caption{Marginal (diagonal) and joint (off-diagonal) posterior distributions for marginal cost parameters $c$ inferred by BNPE from an observed schedule, solution of the stochastic process with nominal parameters $c^\star$. The concentrated, non-uniform posteriors demonstrate successful learning beyond the flat prior (uniform on [10, 50] €/MWh), with nominal cost parameters consistently lying in high-density regions. The inverse optimization solution shows larger deviations, as it assumes deterministic optimality while the generating process is stochastic.}
    \label{fig:posterior_corner_plot_sbi_gen}
\end{figure}
\begin{figure}
    \centering
    \includegraphics[width=\linewidth]{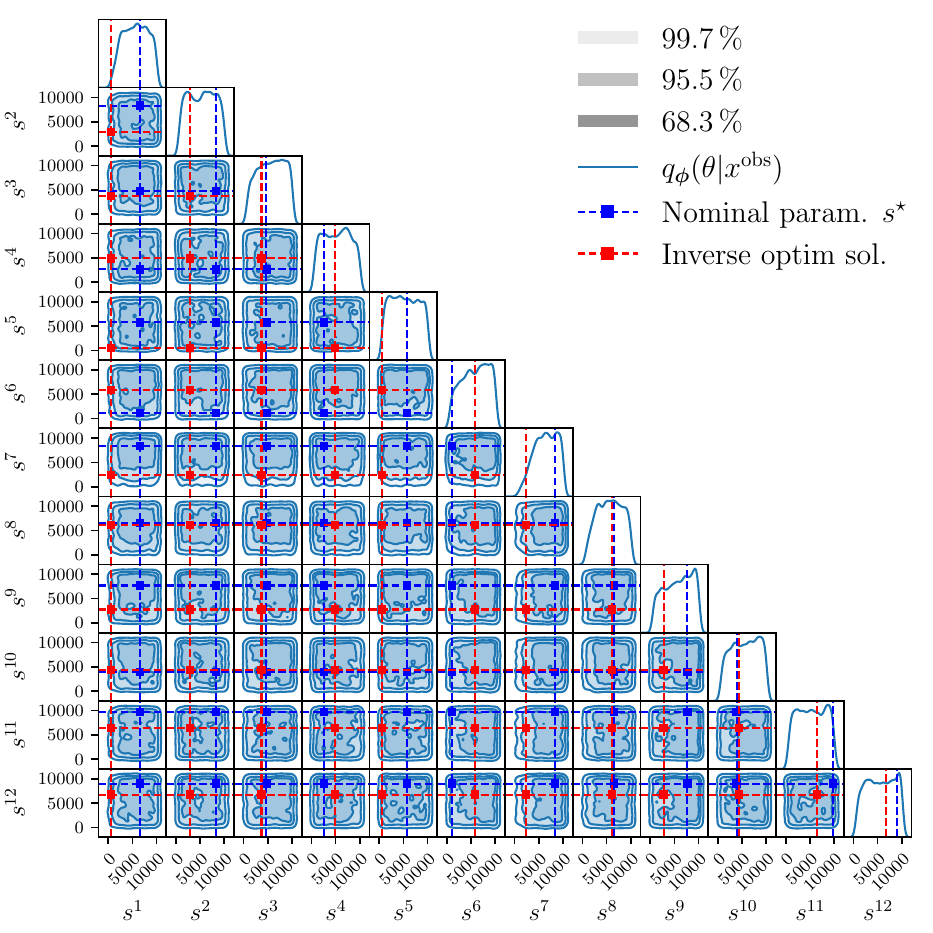}
    \caption{Marginal (diagonal) and joint (off-diagonal) posterior distributions for start-up cost parameters $s$ inferred by BNPE from an observed schedule, solution of the stochastic process with nominal parameters $s^\star$. The nearly uniform posteriors indicate limited learning beyond the prior (uniform on [500, 10000]€), reflecting the challenge of inferring start-up costs from generation schedules. Nominal parameters $s^\star$ often lie in higher-density regions, though posterior concentration remains weak compared to marginal costs.}
    \label{fig:posterior_corner_plot_sbi_start}
\end{figure}

For BNPE training, we generate $N = 2^{18} \approx 262,000$ simulation pairs $(\theta, x)$ where $x=(g, \delta)$ for training and additional $2^{18}$ pairs for validation. The neural density estimator is a neural spline flow \cite{durkan2019neural} with 5 coupling transformations, each parameterized by a masked autoregressive network with 5 hidden layers of 256 units and ReLU activation. All cost parameters, generation schedules, and loads are standardized to zero mean and unit variance before training.

For inverse optimization, we set a maximum number of $100$k iterations and an accuracy threshold at $10e^{-6}$, but the algorithm always achieves this threshold before reaching the maximum number of iterations.

\subsection{Results and diagnostics}

A fundamental challenge in simulation-based inference is that the true posterior distribution $p(\theta|x)$ is unknown; therefore, we cannot assess whether the learned approximation $q_\phi(\theta|x)$ has converged to the target. The quality of the learned posterior depends on the modeling choices that include the prior specification, neural architecture and hyperparameters, the distribution of training data, and the strength of the regularization. To address this lack of ground truth, multiple diagnostic tools can be used to detect potential failures and validate the reliability of these approximations.

A first standard diagnostic is to verify that the posterior conditioned on a given observation does not reproduce the prior and places a high density around the parameter vector that generated the observation. The observed market outcome $x^{\mathrm{obs}}$ is produced in a controlled setup by sampling a nominal parameter vector $\theta^\star$ from the prior and running the forward model. If $q_{\phi}(\theta|x^{\mathrm{obs}})$ matches the prior, it means that the surrogate model does not encode any information about the parameters. Figures~\ref{fig:posterior_corner_plot_sbi_gen} and \ref{fig:posterior_corner_plot_sbi_start} show the distribution of $N = 2^{12}$ samples drawn from the learned posterior $q_\phi(\theta|x^{\mathrm{obs}})$ together with the nominal parameter $\theta^\star$ (marked in blue) and the solution of the inverse optimization problem (marked in red).

For marginal costs (Figure~\ref{fig:posterior_corner_plot_sbi_gen}), the nominal parameter values always lie in high-density regions. The narrow, peaked marginal distributions represent learning beyond the flat uniform prior $\mathcal{U}(10, 50)$ €/MWh, indicating that the posterior reflects information extracted from the observed schedule. Off-diagonal plots show correlations between marginal costs, reflecting how schedules couple the inference of different cost parameters. In contrast, start-up cost posteriors (Figure~\ref{fig:posterior_corner_plot_sbi_start}) remain close to the prior $\mathcal{U}(500, 10000)$€. This diffuse distribution accurately represents the parameter's lack of observability in the forward model; since the schedule $g$ is largely insensitive to $s_j$ once a unit is committed. The fact that the model recovers the prior distribution instead of collapsing to an arbitrary point suggests that BNPE has successfully captured the physics of the problem rather than experiencing a learning failure.

The inverse optimization solution shows larger deviations from the nominal parameters $c^\star$ and often falls into low-density regions of the posterior surrogate. Unlike BNPE, inverse optimization only uses the deterministic part of the model, assuming that the observed schedule $x^{\text{obs}}$ is a perfectly optimal response to the cost parameters $\theta^\star$. When the observation is instead generated by a stochastic process, inverse optimization incorrectly interprets random noise as a change in the UC model.

\begin{figure}
    \centering
    \includegraphics[width=\linewidth]{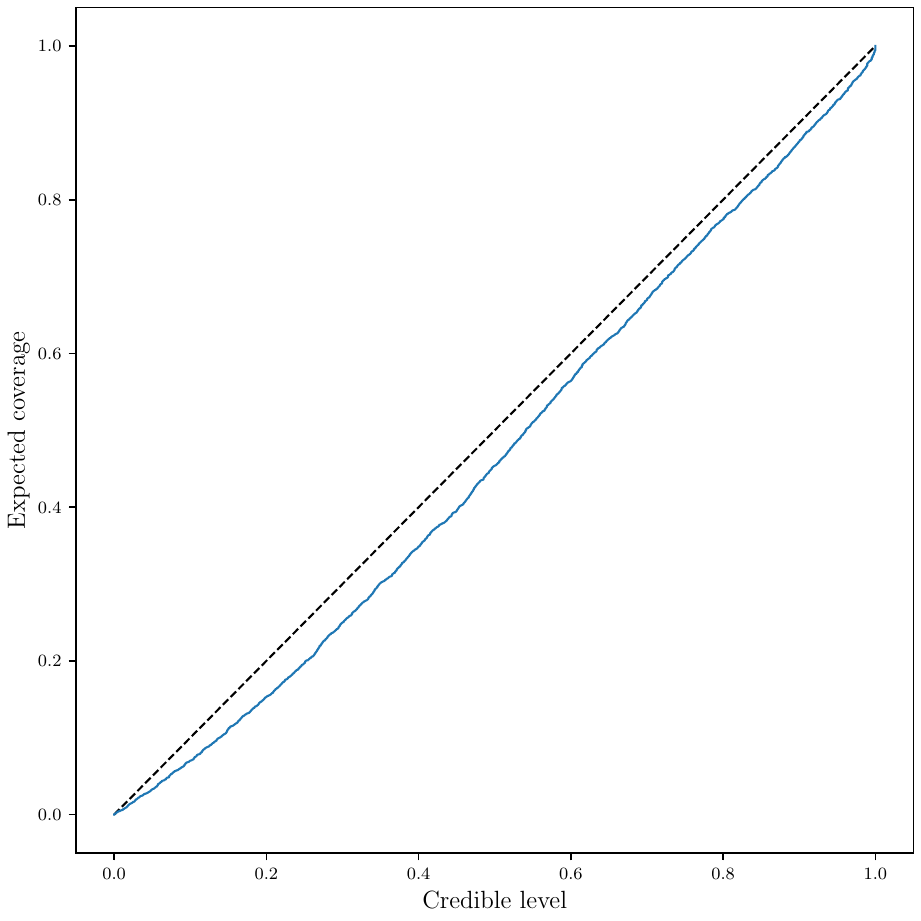}
    \caption{Empirical coverage probability versus nominal credibility level for the learned BNPE posterior. The diagonal line represents perfect calibration, where empirical coverage equals the nominal level $1-\alpha$. The observed curve falls slightly below the diagonal across most credibility levels, indicating modest overconfidence, i.e., credible regions are narrower than ideal. This miscalibration persists despite balanced regularization ($\lambda=50$), likely due to the high-dimensional parameter space (24 parameters) and complex stochastic forward model.}
    \label{fig:coverage_bnpe}
\end{figure}

A second diagnostic assesses whether the surrogate posterior distribution provides the correct level of uncertainty about the parameters that generated an observation. If so, the model is said to be well-calibrated. We validate this by sampling nominal parameters $\theta^\star$ from the prior and generate observations $x^\star$ from the forward model. For a given observation, a $(1-\alpha)$ credible region is defined as a subset of the parameter space containing $(1-\alpha)$ of the posterior mass. For each pair $(\theta^\star, x^\star)$, we construct a $(1-\alpha)$ credible region from $q_{\phi}(\theta|x^\star)$ and check whether $\theta^\star$ lies inside. The expected coverage is the proportion of pairs where the nominal parameter falls within the credible region. A model is well-calibrated if the expected coverage equals $(1-\alpha)$. If the coverage is lower, the model is overconfident, which implies that the learned posterior is too narrow. If the coverage is higher, it is underconfident meaning that credible regions are unnecessarily large. Mathematical details are provided in Appendix~\ref{app:coverage}.



\begin{figure*}[h]
    \centering
    \includegraphics[width=\linewidth]{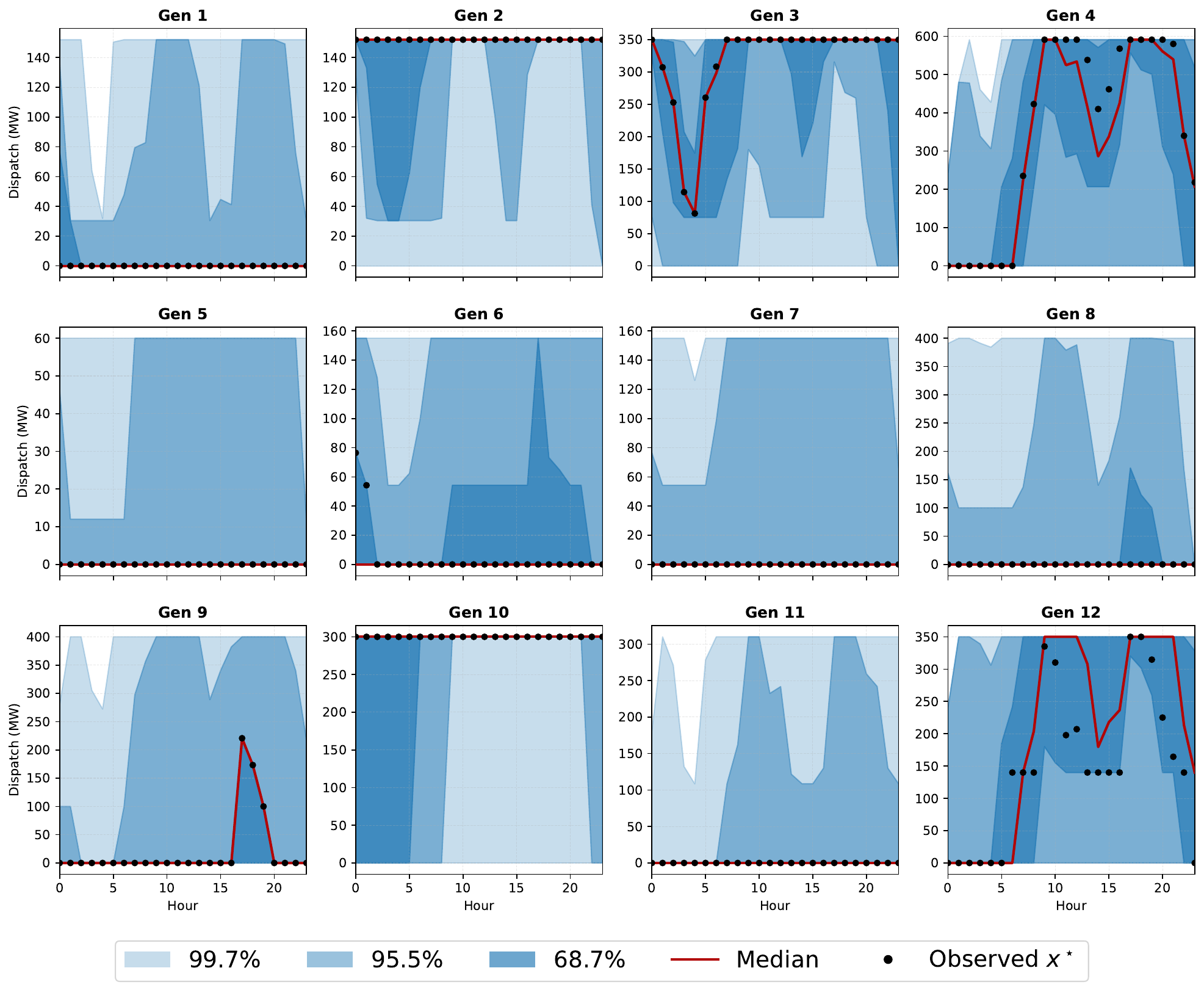}
    \caption{Posterior predictive check. For a single observation $x^\mathrm{obs}$ generated from nominal parameters $\theta^\star$, we sample $N=2^{12}$ parameters from the learned posterior $q_\phi(\theta|x^\mathrm{obs})$ given this observed market outcome and pass them through our forward model. Each subplot shows one generator's dispatch over 24 hours. The observed schedule consistently lies within high-probability regions across all generators and time periods, validating that the learned posterior correctly captures the stochastic forward process.}
    \label{fig:posterior_predictive_check}
\end{figure*}

Our results (Figure~\ref{fig:coverage_bnpe}) show minor overconfidence; empirical coverage falls slightly below the nominal level across all credibility levels. The miscalibration likely arises from the high-dimensional parameter space (24 parameters), the stochastic forward model, and complex interactions between cost parameters and schedules that challenge the neural density estimator's capacity. This uncertainty quantification remains valuable for typical market analysis tasks, where approximate probabilistic bounds inform decision-making even if not perfectly calibrated. In contrast, overconfidence would have been problematic for safety-critical applications where credible intervals directly inform operational margins and underestimating uncertainty could lead to insufficient safety buffers. For electricity market monitoring, the learned posterior provides a reasonable balance between narrowing the prior and avoiding the complete absence of uncertainty inherent to point estimation methods.

A final diagnostic called posterior predictive checks is typically used in real-data applications to assess model adequacy. The idea is to sample parameters from the learned posterior given an observation $q_\phi(\theta|x^\mathrm{obs})$, pass them through the forward model, and check whether the resulting predicted schedules $x^\mathrm{pred}$ are consistent with the observation $x^\mathrm{obs}$. This produces samples from the posterior predictive distribution
$$p(x^\mathrm{pred}|x^\mathrm{obs}) = \int p(x^\mathrm{pred}|\theta) \, q_\phi(\theta|x^\mathrm{obs}) \, d\theta.$$
If the model is misspecified or the posterior is poor, the prediction will deviate from the observation.

In our simulated setting, this diagnostic serves a different purpose. Because the observed market outcome $x^{\mathrm{obs}}$ is itself generated by the same forward model that we use in the predictive check, the goal is not to detect model misspecification but to verify self-consistency. Specifically, we check whether sampling parameters from the learned posterior and pushing them forward through the model reproduces schedules compatible with $x^{\mathrm{obs}}$. If this loop is coherent, then the posterior has successfully captured the inverse mapping.

Figure~\ref{fig:posterior_predictive_check} shows the posterior predictive distributions for all generators over the 24-hour horizon. The learned posterior $q_\phi(\theta|x^\mathrm{obs})$ appears to capture the inverse mapping accurately, as parameters sampled from it generate schedules that are consistent with the observed schedule. This confirms the high quality of the posterior approximation by validating that the entire forward-inverse-forward loop is consistent with the observed data.

\section{Discussion} \label{sec:discussion}

Correctly quantifying model uncertainties is important for avoiding erroneous reasoning and enabling risk-aware decision-making, among other things. As illustrated in the experiments, the most standard approach based on inverse optimization does not quantify uncertainty and suffers from model misspecification; it is therefore prone to such failures. Approximating the posterior distribution may, however, be a difficult task. Markov Chain Monte Carlo (MCMC) methods are widely considered the gold standard for Bayesian inference, as they provide asymptotically exact samples from the posterior distribution under regularity conditions. However, these methods require evaluating the likelihood $p(x|\theta)$. In our setting, the likelihood cannot be computed efficiently. Likelihood-free MCMC variants, such as pseudo-marginal MCMC \cite{andrieu2009pseudo} or ABC-MCMC \cite{marjoram2003markov}, bypass explicit likelihood evaluation by approximating acceptance probabilities through Monte Carlo sampling. However, these methods require solving the UC problem many times per MCMC step. Additionally, these methods suffer exponentially vanishing acceptance rates in high dimensions. Together, these limitations make likelihood-free MCMC computationally prohibitive for this application.

Amortized optimization methods and in particular simulation-based inference (SBI) partially address previous limitations. SBI requires a one-time training phase but then provides fast posterior evaluation for any new observation, making it ideal for operational use where the same model is inverted repeatedly (e.g., daily cost estimation). For applications making many inferences between model updates, SBI's retraining cost is small compared to the cumulative cost of running inverse optimization or MCMC for each query. In contrast, inverse optimization and MCMC adapt immediately without requiring retraining when the model changes because of regulatory updates, network modifications, or improved modeling. This makes them better suited for research settings with evolving infrastructure.

\section{Conclusion} \label{sec:conclusion}
We formulated generation cost estimation from market data as a Bayesian inference problem with a latent variable model accounting for opportunity costs and operational uncertainty. The Bayesian approach allows for encoding prior knowledge about cost parameters and systematically updating beliefs given observed data. Two complementary inference methods were discussed. Balanced neural posterior estimation (BNPE) that uses the full latent variable model to learn amortized posterior approximations with full uncertainty quantification. Second, feasibility-based inverse optimization using only a unit commitment problem provides fast point estimates suitable for time-critical decisions.

Empirical validation on the IEEE RTS-96 test system demonstrates successful parameter estimation. Marginal costs are accurately inferred with concentrated posteriors, while start-up cost posteriors appropriately remain diffuse, correctly reflecting their limited observability from schedule observations alone. BNPE provides well-calibrated approximate posteriors with fast amortized inference, meaning that once trained, posterior evaluation for new observations is nearly instantaneous, enabling real-time deployment for operational market monitoring. This approach successfully combines uncertainty quantification with computational efficiency, offering a practical solution for cost parameter estimation in electricity markets.

Several promising directions extend this work. First, inferring physical parameters (e.g., ramp rates, capacity limits) jointly with cost parameters would account for model misspecification by modeling all uncertain quantities as random variables. For high-dimensional parameter spaces exceeding tens of parameters, recent advances in flow matching \cite{dax2023flow} offer more scalable alternatives to normalizing flows used here. Second, extending inference across multiple days could strengthen identification since generator marginal costs vary with publicly available fuel prices (gas, coal, oil). This would substantially increase the effective sample size, tightening the posterior and potentially identifying start-up costs from temporal patterns in consecutive start-up decisions. Finally, while Gaussian noise currently approximates unknown opportunity costs, future work should incorporate more complex distributions to better capture real-world trading strategies.

\newpage

\IEEEtriggeratref{19}


%
\bibliographystyle{IEEEtran}
\bibliography{bibliography.bib}

\newpage
\onecolumn
\section{Appendix}

\subsection{Derivation of the BNPE objective}
\label{app:bnpe_derivation}

We derived the BNPE objective by minimizing the expected Kullback-Leibler (KL) divergence between the true posterior $p(\theta \mid x)$ and its approximation $q_\phi(\theta \mid x)$
\begin{equation}
  \label{eq:bnpe_objective}
  \min_{\phi} \underset{p(x)}{\mathbb{E}} \left[ \mathrm{KL} \big( p(\theta \mid x) \parallel q_\phi(\theta \mid x) \big) \right].
\end{equation}
The KL divergence is defined as
\begin{align}
    \mathrm{KL} \big( p(\theta \mid x) \parallel q_\phi(\theta \mid x) \big) = &\int p(\theta \mid x) \log \frac{p(\theta \mid x)}{q_\phi(\theta \mid x)} \, d\theta, \notag\\
    = &\int p(\theta \mid x) \big[ \log p(\theta \mid x) - \log q_\phi(\theta \mid x) \big] \, d\theta. \notag\\
\end{align}
Using Bayes' rule, $p(\theta \mid x) = \frac{p(\theta) p(x \mid \theta)}{p(x)}$
\begin{align}
    &\int p(\theta \mid x) \big[ \log p(\theta \mid x) - \log q_\phi(\theta \mid x) \big] \, d\theta, \notag\\
    = &\int \frac{p(x \mid \theta) \; p(\theta)}{p(x)} \big[ \log p(\theta \mid x) - \log q_\phi(\theta \mid x) \big] \, d\theta. \notag\\
\end{align}

The full expectation becomes
\begin{align}
\underset{{p(x)}}{\mathbb{E}} \left[ \mathrm{KL} \big( p(\theta \mid x) \parallel q_\phi(\theta \mid x) \big) \right] &= \int p(x)\int \frac{p(x \mid \theta) \; p(\theta)}{p(x)} \big[ \log p(\theta \mid x) - \log q_\phi(\theta \mid x) \big] \, d\theta \; dx, \notag\\
&= \int \int p(x \mid \theta) \; p(\theta) \big[ \log p(\theta \mid x) - \log q_\phi(\theta \mid x) \big] \, d\theta \; dx, \notag\\
&= \underset{{p(\theta) \, p(x \mid \theta)}}{\mathbb{E}} \left[ \log p(\theta \mid x) - \log q_\phi(\theta \mid x) \right]. \notag
\end{align}

Minimizing this is equivalent to maximizing the expected log-probability of the approximate posterior
$$ \underset{p(\theta) \, p(x \mid \theta)}{\mathbb{E}} \left[ \log q_\phi(\theta \mid x) \right],$$
since the entropy term $\log p(\theta \mid x)$ is independent of the network parameters $\phi$.

\subsection{Coverage}
\label{app:coverage}

Let ${\boldsymbol{\Theta}}_{q_{\phi}(\theta|x)}(1-\alpha)$ be the set of regions in the parameter space containing at least $1-\alpha$ probability mass of the learned posterior $q_{\phi}(\theta|x)$

$${\boldsymbol{\Theta}}_{q_{\phi}(\theta|x)}(1-\alpha) = \left\{
\Omega \; \middle| \;
\int_{\Omega} q_\phi(\theta|x) \, d\theta \ge 1-\alpha
\right\}.$$

Let ${\boldsymbol{\Theta}}^\star_{q_{\phi}(\theta|x)}(1-\alpha) \in {\boldsymbol{\Theta}}_{q_{\phi}(\theta|x)}(1-\alpha)$ be the $(1-\alpha)$-highest posterior density region

$$ {\boldsymbol{\Theta}}^\star_{q_{\phi}(\theta|x)}(1-\alpha)  = \underset{\Omega \in {\boldsymbol{\Theta}}_{q_{\phi}(\theta|x)}(1-\alpha)}{\arg \sup} \quad \inf_{\theta \in \Omega} \, q_\phi(\theta|x).$$

The expected coverage is
\begin{equation}
    \text{Cov}(1-\alpha) = \underset{p(\theta, x)}{\mathbb{E}} \big[ \mathds{1}\{\theta \in {\boldsymbol{\Theta}}^\star_{q_{\phi}(\theta|x)}(1-\alpha) \} \big].
\end{equation}

In practice, we estimate this empirically on a test set of $N=2^{12}$ samples $(\theta^{(i)}, x^{(i)})$ by
\begin{equation}
    \widehat{\text{Cov}}(1-\alpha) = \frac{1}{N}\sum_{i=1}^N \mathds{1}\{\theta^{(i)} \in {\boldsymbol{\Theta}}^\star_{q_{\phi}(\theta|x^{(i)})}(1-\alpha) \}.
\end{equation}

Perfect calibration yields $\widehat{\text{Cov}}(1-\alpha) = 1-\alpha$ for all credibility levels $(1-\alpha)$.

\end{document}

%% file: Authors.tex
\author{

    \IEEEauthorblockN{
        Matthias Pirlet$^{a*}$,
        Adrien Bolland$^a$,
        Alexandre Huynen$^b$,
        Quentin Louveaux$^a$,
        Gilles Louppe$^a$,
        Damien Ernst$^a$
    }
    
    \IEEEauthorblockA{
        $^a$University of Liège, Belgium; \,
        $^b$Engie, Brussels, Belgium
    }

    \IEEEauthorblockA{* matthias.pirlet@uliege.be}
}

\maketitle